# Privacy Risk and Preservation For COVID-19 Contact Tracing Apps[1]

**Dong Wang**

State Key Laboratory of Information Engineering in Surveying,

Mapping and Remote Sensing, Wuhan University, China, 430079

**Fang Liu**

Applied & Computational Mathematics and Statistics,

University of Notre Dame, USA, 46530

fliu2@nd.edu

The WHO declared the coronavirus COVID-19 outbreak a public health emergency of international concern on January 30, 2020. Since the first COVID-19 case on November 17, 2019 (according to unpublished government data), the number of cumulative cases worldwide has been around 9.76 million and more than 492 K had died from the disease as of June 27, 2020.

A huge amount of data have been and are being collected during the pandemic, and will be in the future. The data, coupled with state-of-art computing and analysis techniques, play a powerful role in the efforts of harnessing the spread of COVID-19. On the other hand, the collected data, such as health data and medical history, contact tracing, and social control, often contain personally identifiable information and pose high risk for compromised privacy. The EU Parliament states in a press release that "These tools could seriously interfere with people's fundamental rights to a private life and the protection of personal data, and are tantamount to a state of surveillance of individuals". How to balance private protection vs personal data collection and release for monitoring the pandemic and improving public health has attracted much research interest and will be a subject of continuing debate. Government, academia, and industry have already worked together to search for effective solutions to this problem.

A variety of types of information collected during the COVID-19 pandemic are associated with privacy concerns. Here, we focus on the privacy issues incurred by contact.

**Contact Tracing in COVID-19 Pandemic**

Contact tracing is to use digital tools to trace and monitor contacts of infected people during an epidemic, so to alert and inform people who have come into contact with interacted people, and help to ensure effective quarantine of contacts to prevent additional transmission. Contact tracing is widely adopted and is key to prevent the further spread of COVID-19 in many regions and countries. Figure 1 was used by Washington Governor Jay Inslee when announcing the state's contact tracing plan and shows a nice summary on how contact is used and can do.

---

[1] To appear in CHANCE 33(3): Special Issue on Covid-19.

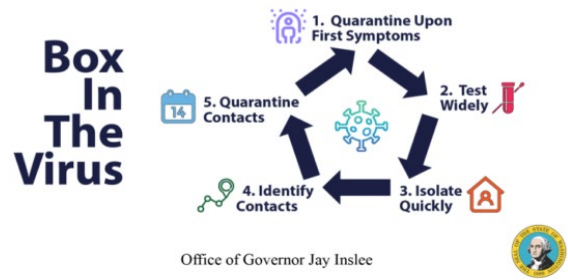

Figure 1: Contact Tracing (courtesy of Washington Governor Office)

Many countries and regions have the ability to conduct manual contact tracing; however, it is difficult to scale up manual tracing to respond to the unpreceded COVID-19 pandemic, not to mention the resources spent on training human contact tracers and that the tracing process is prone to time-consuming and error-prone. To help (semi-)automatize the contact tracing and notification efforts, countries and regions have developed and deployed or are considering adopting contact-tracing software or mobile apps. While contact tracing apps and software are playing an important role in tracking and slowing down the spread of COVID-19, red flags have been raised regarding the high privacy risk associated with contact tracing in the media such as [Washington Post](), [Forbes,]() and [Reuters](), as well as [organizations]() and [government agencies](). For example, contact tracing based on location data often collects very detailed and frequent location data, which are known to be highly revealing of people's identity and can be used to infer of the private social life and health status of the individuals (De Montjoye et. al. in 2013 published a study of 1.5 million individuals over a period of 15 months found that four spatial-temporal mobility data points are enough to identify 95% of the individuals).

Table 1 provides some examples on the contact tracing apps and software in some countries that categorized by the technology used and the degree to which authorities are involved during data collection and information sharing. GPS-based contact-tracing apps and software collect users' location data, whereas the Bluetooth based techniques mostly only require the relative tempo-spatial proximity among users. In that sense, less private information is collected in the Bluetooth-based approaches than in the GPS-based approaches. In both the GPS and the Bluetooth-based approaches, the centralized or non-centralized models can be deployed to collect and store data, share information, and alert users regarding potential COVID-19 exposure. However, the two models differ in the levels of anonymity and in the approaches to achieve privacy protection for the data contributors. For the centralized model, contact tracing data are collected, integrated, shared with targeted individuals by some authorities (e.g health authorities or federal, state or local governments). Therefore, the centralized model somewhat operates like a mass surveillance system and data are collected from everyone, whether healthy or diseased, and the authorities have the unique identifiers on all the individuals and know whom to target with certain information. In terms of privacy, there is no privacy for the users in terms information sharing with the health authorities and they will just have to trust the authorities to keep their data safe and private. In contrast, for the decentralized model, there is no need to collect or store information on everyone through a central server. Location and contact information of those who are not tested or tested negative are stored and processed locally on their respective devices, and they can choose to check whether they have crossed path with infected people through public platforms like a website the contains COVID-19 hot spots information built by authorities or receive notification by locally information matching. In the former case, the information shared on the public website has already gone through some types of data anonymization, blurring, and redaction through careful planning or integrating formal privacy concepts. In summary, the decentralized model offers a higher level of privacy protection on individuals compared to the centralized models. Table 1 also suggest the centralized model is mainly employed by Asian countries, whereas the decentralized model is preferred by the US and European countries. In what follows, we will look into how the centralized and decentralized models work in the GPS and Bluetooth based technology, respectively, in more details.

Table 1: Examples of COVID-19 Contact Tracing Apps and Software

| | | technology | | |
|---|---|---|---|---|
| | | GPS | Bluetooth | GPS+Bluetooth |
| **data collection and information sharing model** | centralized | Alipay Health Code (China); WeChat (China) ; Corona 100m (South Korea); CovidTracker (Thailand); ProteGo (Poland) | Tracetogether (Singapore) | Aarogya Setu (India) |
| | decentralized | safe paths (US) HaMagen (Israeli) | Pan-European Privacy-Preserving Proximity Tracing (PEPP-PT) (EU); COVID watch (Mainly US); PACT (US); COVIDsafe (Australia) | |

## GPS-based Contact-Tracing Scheme

GPS-based apps collect time-stamped GPS points 24/7 from individuals. If the collected GPS data suggest that two people were in close proximity to each other at a certain time, and if one of them was tested positive for COVID-19 later on, then the other person will either receive notifications from authorities regarding the contact event or find out this information by locally checking publicly posted contact tracing information from the health authorities themselves. The way that the authorities collect and share information leads to either the centralized and decentralized models.

[The Alipay Health Code from China](#) is an example of the centralized model. Alipay Health Code assigns an individual a color QR code (green, yellow, red), representing the individual's health status. A green code indicates the highest of level of healthiness and the individual is allowed to go anywhere unrestricted; red stands for high risk and requires a 2-week quarantine, and yellow means a 1-week quarantine. The determination of the color code is often based on the location history of the individual. If the person has been to a COVID-19 hot spot, then there is a non-ignorable chance the person may be infected and is likely to receive a red or yellow code. Each time the individual's QR code is scanned, the information regarding the current location is sent to some servers belonging to some authorities, allowing the authorities to track people's movements over time. Furthermore, the app often requires users to register with their unique identification information such as national identification number, name, and phone number. Similar to China, South Korea developed the [Corona 100m (Co100) app](#) in the framework of the centralized model. The app uses the government collected location data to alert users when they come within 100 meters of a location recently visited by a COVID-19 patient.

[Safe Paths](#) is an MIT-led privacy-preserving platform and is an example of the decentralized model. It comprises a smartphone app, PrivateKit, and a web application, SafePlaces. SafePlaces share anonymized and blurred location histories of infected people whereas PrivateKit allows users to match their personal location history with shared information on SafePlaces. In other words, healthy individuals keep their own location diaries without having to sharing with or reporting to authorities. If an individual is tested positive for COVID-19, her location history information, with her consent, will be reported to the authorities. Since the location history contains private information and a big part of it is not even relevant for COVID-19 tracing (e.g. the home location where the individual spends most of her time), the information is often redacted or blurred before placed on SafePlaces, to which users can compare their location diaries with those infected to see if they have ever crossed paths. Israel developed the [Hamagen app](#) (Hamagen is Hebrew for shield) that is based on a similar idea as the Safe Paths platform. It allows local comparison of users' GPS data with the government epidemiological location database on COVID-19 hot spots.

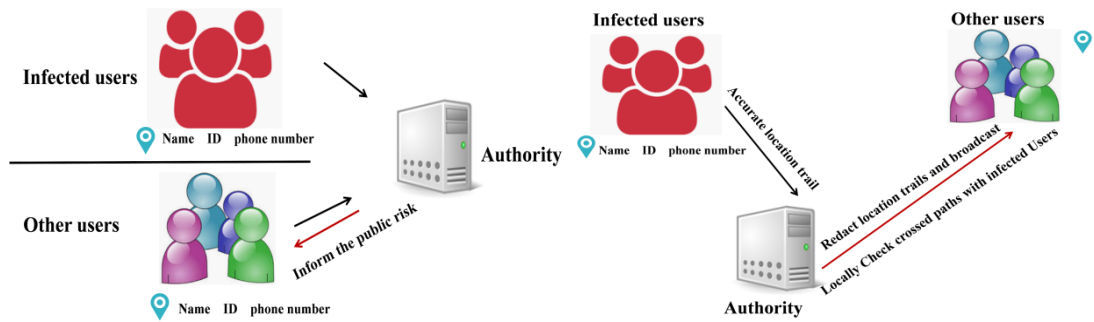

(a) A schematic of GPS-based centralized model    (b) A schematic of GPS-based decentralized model

Figure 2: GPS-based contact-tracing schemes

The centralized and decentralized models in the framework of GPS-based contact-tracing are depicted in Figure 2. The centralized model tracks location and contact information and health status with unique identifiers from both patients and healthy users. The potential privacy risk in this scheme is obvious. First, it can be tricky to keep the identity of the infected person confidential in some cases when broadcasting her location history, especially if she is one of the few persons the healthy people interacted with in close proximity recently. Second, it personalizes the alert and notification systems. This level of precision comes at the cost of compromised individual privacy. Individuals whose data are collected in the centralized model reply on the authorities to keep their data private and safe, but this is not always warranted. In contrast, the decentralized model only collects location information and shares an anonymized version of that information from reported COVID-19 patients. Therefore, there is a higher level of privacy protection for the patients. Furthermore, the decentralized model does not trace healthy people and only share the information on COVID-19 hot spots on a public forum with no specific targeting at certain individuals or sending un-personalized information to app users. Therefore, the privacy concern level is much lower in the decentralized model for healthy people. On the other hand, without personalized alerts, the decentralized model would reply on users' self-initiation and pro-activity to check whether they might have been to any of the infection hot spots recently and self-quarantine if that is the case.

Formal notions on privacy guarantee can be incorporated in both the centralized and decentralized models when developing the GPS-based contract tracing apps and software. For example, the $k$-anonymity model introduced by Sweeney in 2002 can be used to collapse detailed location information or quasi-identifiers to yield at $\geq k$ "homogeneous" individuals in each cell of the cross-tabulation of a set of attributes. Geo-indistinguishability is a formal location privacy concept proposed by Andrés et. al. in 2013, extending of the popular differential privacy concept by Dwork et. al. in 2006, and can be used to generate sanitized location information. Regardless which formal privacy notation is used to sanitize the data before releasing and sharing with the public or a targeted group, the accuracy of contact tracing will be more or less affected.

# Bluetooth-based Contract-Tracing Scheme

Different with the GPS -based privacy-preserving scheme, the Bluetooth-based contact-tracing apps do not collect the exact location information from their users. In that sense, users would feel more private and less anxious about being monitored 24/7 on their whereabouts. In addition, Bluetooth has higher contact tracing accuracy than GPS-based apps. Bluetooth signals do not rebound and pass through most soft walls, helping to avoid the false positive that two people in close proximity are regarded as a "contact" event whereas they are actually separated by walls.

Specifically, the Bluetooth technology is leveraged to collect information on whether or not two people have appeared in the same location within 6 feet of each other at the same time. Each app user generates a time-varying sequence of random tokens, which are stored locally on their devices. The time interval between two tokens cannot be too frequent to cause computational or storage difficulty for the users, and nor can be too infrequent to make the tracing ineffective or incur privacy concerns. If two users appear within 6 feet of each other at a time $t$, they exchange their tokens at that time, which are stored in their contact token sets. If one of them is diagnosed with COVID-19, say within 2 weeks after the contact event, then she will, with her consent, share her contact token sets from the last 2 weeks with health authorities who will subsequently notify people who might have had shared a contact event with the infected person. The way that the authorities collect data and notify users about potential exposure leads to either the centralized or de-centralized models.

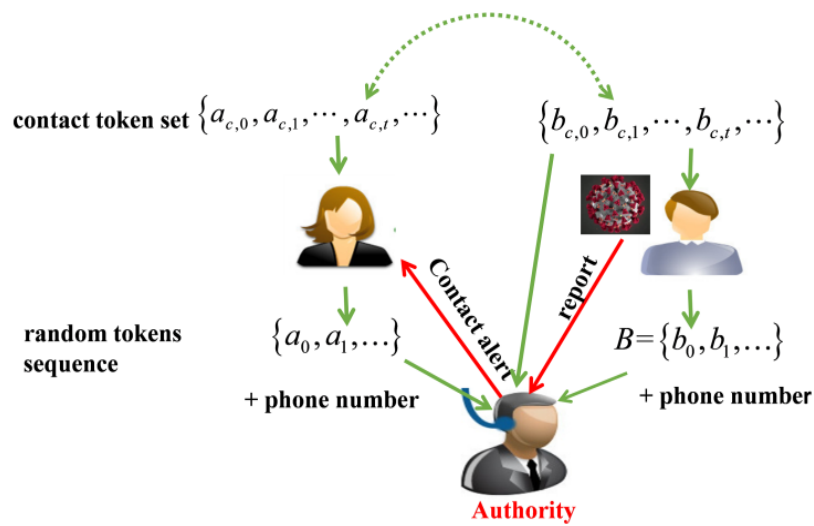

Figure 3: A schematic of Bluetooth-based Centralized Model

The Singapore's Bluetooth based mobile phone app TraceTogether is an example of the centralized approach. In this model (Figure 3), all app users report their tokens (denoted by $\{a_0, a_1, \ldots\}$ and $\{b_0, b_1, \ldots\}$) as well as their phone numbers to the health authority regardless of their health status. If a person is diagnosed with COVID-19, the person updates the authority on his health status and shares his contact tokens (denoted by $\{b_{c,0}, b_{c,1}, \ldots, b_{c,t}, \ldots\}$). The authorities then match each token in the contact token set with their database of tokens, and alert the users with matches through their phone contacts.

The privacy risk on the infected individuals regarding their health status associated with the alerting and notification system is similar to the centralized model in the GPS-based scheme. Similar to the centralized model in the GPS-based system, the users have no choice but just have to trust the authorized safe and private. In addition, the authorities have each user' phone number, which is a unique identifier and can be used to link to other databases that might contain sensitive information on the users if the authorities feel there is a need to do so.

The [Covid Watch](), Private Automated Contact Tracing ([PACT]()), and [COVIDsafe]() apps and the [Pan-European Privacy-Preserving Proximity Tracing (PEPP-PT)]() software are examples of the decentralized approaches that leverage the Bluetooth technology. The Covid Watch app represents an international effort from more than 400 volunteers around the world (US, Canada, Australia, etc), and sends anonymous privacy-preserving COVID-19 exposure alerts via private and local Bluetooth signal. PACT is developed by MIT, working with partners from around the world; not only collect information on binary contact events but also on the distance and time duration of a contact event. COVIDsafe is the app used by Australian government for contact tracing. PEPP-PT is a larger software system with many individual components and sends alert on possible exposures. In the decentralized model (Figure 4), only the tokens of the patient with COVID-19, together with his history contact event numbers, are collected by authority, with the patient's consent. The patient receives a permission number from the authority, which is then posted in a public database. The public database verifies the permission number and updates itself with the contact numbers shared by the patient. Other users can compare their contact event numbers with the publicly posted contact event numbers; and if there is a match, they are noticed that they may have been exposed to the virus and need self-quarantine.

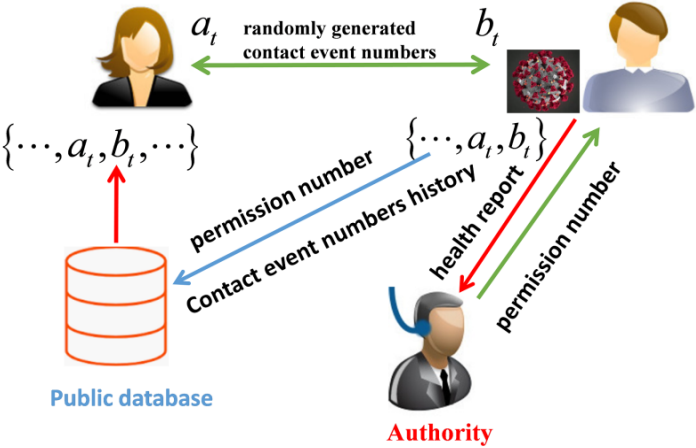

Figure 4: A schematic of Bluetooth-based Decentralized Model

For the Bluetooth-based privacy-preserving contract tracing, anonymity can be enhanced by randomly swapping generated random tokens across users to better prevent linkage privacy attack. There also exist cryptography solutions toward privacy protection, such as the technology developed by [Apple and Google](#), using secure multi-party computation without relying on a trusted server, or sending anonymous encrypted or random messages, as proposed by Cho et. al, Hekmati et. al, and Reichert et al. in March and April of 2020.

## Final Remarks

In addition to the technology-specific approaches for privacy-preserving contact tracing, some general principles for data privacy protection, such as necessity, proportionality, and transparency, may also apply in the COVID-19 pandemic data collection and information sharing. For example, it is often permissible to share anonymized data or aggregated statistics that are associated with low individual re-identification risk. If there exists non-ignorable re-identification risk or there is a need to reveal individual identity when releasing information, there must be a justification for doing so. Minimizing data collection and limiting access and retaining data only for the minimum amount of time that is necessary also help to reduce the privacy harms due to COVID-19 data processing. Obtaining consents is also commonly used for privacy protection. The subjects from whom data are collected and shared should receive clear communications from authorities regarding the purposes and usage, and the retention duration of their data, among others. Given the unprecedented situation of COVID-19, consents might take a different form than the regular consents when it comes to data sharing, especially when an individual feels compelled to share her contact and location history once tested positive.

Privacy is not at odds with the efforts and measures for harnessing COVID-19. On the contrary, any measure taken authorities and governments is effective only when its citizens trust such a measure. It is pointless to develop a contact-tracing software or app that users don't trust or use. The protection of privacy is key to building such trust between the public and the authorities in the case of COVID-19 contact tracing. Government and authorities should continue to commit to privacy preservation during their push and implementation of COVID-19 contract tracing, now and in the future. All parties (the public, authorities, academic, and industry) should work together to develop effective policies and technologies to protect the privacy of the people when collecting data on COVID-19 to help curb the global pandemic.

## Further Reading

**Covid 19-Pademic**

WHO declared the coronavirus disease 2019. https://www.who.int/emergencies/diseases/novel-coronavirus-2019

First covid-19 case happened in November, china government records show-report. https://www.theguardian.com/world/2020/mar/13/first-covid-19-case-happened-in-november-china-government-records-show-report

COVID-19 pandemic. https://en.wikipedia.org/wiki/COVID-19_pandemic

**Formal Privacy Concepts**

C. Dwork, F. McSherry, K. Nissim, and A. Smith, "Calibrating noise to sensitivity in private data analysis," in Theory of cryptography conference, 2006, pp. 265-284.

M. E. Andrés, N. E. Bordenabe, K. Chatzikokolakis, and C. Palamidessi, "Geo-indistinguishability: Differential privacy for location-based systems," in Proceedings of the 2013 ACM SIGSAC conference on Computer & communications security, 2013, pp. 901-914.

L. Sweeney, "k-anonymity: A model for protecting privacy," International Journal of Uncertainty, Fuzziness and Knowledge-Based Systems, 2002, vol. 10, pp. 557-570.

**Privacy on Location Data and Contact Tracing**

Y.-A. De Montjoye, C. A. Hidalgo, M. Verleysen, and V. D. Blondel, Unique in the crowd: The privacy bounds of human mobility, Scientific reports, 2013, 3:1376.

Contact tracing apps can help stop coronavirus. But they can hurt privacy. https://www.washingtonpost.com/politics/2020/04/28/contact-tracing-apps-can-help-stop-coronavirus-they-can-hurt-privacy/

Coronavirus Contact-Tracing Apps Miss The Point About Privacy https://www.forbes.com/sites/simonchandler/2020/05/04/coronavirus-contact-tracing-apps-miss-the-point-about-privacy/#630d1b2539c4

UK privacy advocates warn over covid-19 contact tracing app. https://www.reuters.com/article/us-health-coronavirus-britain-apps/uk-privacy-advocates-warn-over-covid-19-contact-tracing-app-idUSKBN22C2K0

Use of smartphone data to manage covid-19 must respect EU data protection rules. https://www.europarl.europa.eu/news/en/press-room/20200406IPR76604/use-of-smartphone-data-to-manage-covid-19-must-respect-eu-data-protection-rules

IAPP COVID-19 Guidance and Resources. https://iapp.org/resources/topics/covid-19-guidance-and-resources/

**GPS-based Contact Tracing of Covid-19**

What the US can learn from other countries using phones to track covid-19. https://www.vox.com/recode/2020/4/18/21224178/covid-19-tech-tracking-phones-china-singapore-taiwan-korea-google-apple-contact-tracing-digital

In coronavirus fight, china gives citizens a color code, with red flags. https://www.nytimes.com/2020/03/01/business/china-coronavirus-surveillance.html

South korea to step-up online coron-avirus tracking. https://www.smartcitiesworld.net/news/news/south-korea-to-step-up-online-coronavirus-tracking-5109

Covid-19: Poland launches an official tracking app. https://www.natlawreview.com/article/covid-19-poland-launches-official-tracking-app

Covid-19 news tracker-location-based news about covid-19 in Thailand. https://covidtracker.5lab.co/zh-hans?fbclid=IwAR1bAH4qDAZtWkdh2MVwAiFmow9lAtRFg78vPSZKr76__ezADDlBNwYHTyk

HaMagen- The Ministry of Health App for Fighting the Spread, https://govextra.gov.il/ministry-of-health/hamagen-app/download-en/

N. Sawant. (2020, 5/17). Aarogya setu: Whether we like it or not, the app is here to stay, but it's still riddled with privacy issues that need strong answers. https://www.firstpost.com/tech/news-analysis/aarogya-setu-whether-we-like-it-or-not-the-app-is-here-to-stay-but-its-still-riddled-with-privacy-issues-that-need-strong-answers-8348131.html

Covid watch. https://www.covid-watch.org/

Safe paths. http://safepaths.mit.edu/

**Bluetooth-based Contact Tracing of Covid-19**

Tracetogether, safer together.  https://www.tracetogether.gov.sg/

Pact: Private automated contact tracing.  https://pact.mit.edu/

Pan-European Privacy-Preserving Proximity Tracing (PEPP-PT) (EU), https://www.pepp-pt.org/

Australian Government Department of Health COVIDSafe app. https://www.health.gov.au/resources/apps-and-tools/covidsafe-app

Aarogya setu: Whether we like it or not, the app is here to stay, but it's still riddled with privacy issues that need strong answers.  https://www.firstpost.com/tech/news-analysis/aarogya-setu-whether-we-like-it-or-not-the-app-is-here-to-stay-but-its-still-riddled-with-privacy-issues-that-need-strong-answers-8348131.html

Privacy-preserving contact tracing.  https://www.apple.com/covid19/contacttracing

A. Hekmati, G. Ramachandran, and B. Krishnamachari, "CONTAIN: privacy-oriented contact tracing protocols for epidemics", arXiv preprint 2020, arXiv:2004.05251

H. Cho, D. Ippolito, and Y. W. Yu, "Contact tracing mobile apps for COVID-19: Privacy considerations and related trade-offs," arXiv preprint 2020, arXiv:2003.11511

L. Reichert, S. Brack, and B. Scheuermann, "Privacy-preserving contact tracing of covid-19 patients," 2020, https://eprint.iacr.org/2020/375.pdf

## About the Authors

**Dong Wang** is currently working toward the Ph.D. degree in Computer Science at the State Key Laboratory of Engineering in Surveying, Mapping and Remote Sensing of Wuhan University, China. She is currently a visiting Ph.D. student in the Department of Applied and Computational Mathematics and Statistics at the university of Note Dame. Her research interests include data privacy and data mining. She would like to thank the China Scholarships Council program (NO. 201906270230) for supporting her work.

**Fang Liu** is a Professor in the Department of Applied and Computational Mathematics and Statistics at the University of Notre Dame, Notre Dame, IN. Prof. Liu obtained her Ph.D. in Biostatistics from the University of Michigan, Ann Arbor. Her research interests include data privacy and differential privacy, statistical machine learning of big data, model regularization, Bayesian methodology, and analysis of missing data. She would like to thank NSF (Grant #1717417) for supporting her work on data privacy.